\documentclass[aps,prl,twocolumn,floatfix,showpacs]{revtex4}
\newcommand{\be}{\begin{equation}}
\newcommand{\ee}{\end{equation}}
\newcommand{\bea}{\begin{eqnarray}}
\newcommand{\eea}{\end{eqnarray}}

\begin{document}
\title{Magnetic control of the interaction in ultracold K-Rb mixtures}

\author{A. Simoni}
\affiliation{
INFM and Department of Chemistry, University of Perugia, I-06123 Perugia
, Italy}
\author{F. Ferlaino, G. Roati, G. Modugno and M. Inguscio}
\affiliation{
INFM and LENS, Universit\`a di Firenze, Via Nello Carrara 1, 50019
Sesto Fiorentino, Italy }

\begin{abstract}
We predict the presence of several magnetic Feshbach resonances in
selected Zeeman sublevels of the isotopic pairs $^{40}$K-$^{87}$Rb and
$^{41}$K-$^{87}$Rb at magnetic fields up to $10^3$~G
(1G~=~$10^{-4}$~T).  Positions and widths are determined combining a
new measurement of the $^{40}$K-$^{87}$Rb inelastic cross section with
recent experimental results on both isotopes.  The possibility of
driving a K-Rb mixtures from the weak to the strong interacting regime
tuning the applied field should allow to achieve the optimal conditions for
boson-induced Cooper pairing in a multi component $^{40}$K-$^{87}$Rb
atomic gas and for the production of ultracold polar molecules.
\end{abstract}

\maketitle

The possibility of controlling the interactions in an ultracold atomic
gas using magnetically tunable Feshbach
resonances~\cite{inouye98nature} has been successfully exploited in
recent remarkable experiments. It has indeed allowed to reach
Bose-Einstein condensation (BEC) for species difficult to condense in
zero field and to study their stability
properties~\cite{roberts01prl,weber02science}, to form bright
solitons~\cite{khajkovich02science}, to produce ultracold homonuclear
molecules~\cite{donley02nature} and a strongly interacting Fermi
gas~\cite{ohara02science}. 
The presence of Feshbach resonances between atoms of different species
could open new possibilities, thereby in this paper we carry out a
detailed investigation in this direction.
\newline \indent
We focus on two potassium-rubidium mixtures that have
allowed the production of a two-species BEC
($^{41}$K-$^{87}$Rb)~\cite{science1,modugno02prl} and of a strongly
interacting Fermi-Bose mixture ($^{40}$K-$^{87}$Rb)~\cite{roati02prl}.
We predict the positions and the widths of several Feshbach resonances
in selected magnetic sublevels of the ground electronic state by
combining collisional measurements on the two isotopes. These resonances
might for instance allow the observation of boson-induced Cooper
pairing~\cite{bijlsma00pra} or the formation of ultracold bosonic or
fermionic polar molecules~\cite{paul}. 
\newline \indent
We use for our calculations the
collision model presented in Ref.~\cite{ferrari}, where the Hamiltonian
now includes the Zeeman interaction with the external magnetic field~\cite{pot}.
Denoting ``a'' and ``b'' potassium and rubidium, collisions are labeled as
$(f_{a} m_{fa})$+$(f_b m_{fb})$, where $f_a$, $f_b$ are the hyperfine
angular momenta with projection $m_{fa}$, $m_{fb}$ along the
quantization axis. Please recall that the model is parameterized by the
singlet and triplet $s$-wave scattering lengths $a_{s,t}$ and by the
long-range dispersion coefficients $C_n$ with $n$=6,8,10. We adopt a
$\pm2\%$ uncertainty in the highly accurate van der Waals coefficient
$C_6=4274$~a.u.  (1~a.u.~=~0.0957345$\times$10$^{-24}$J$\cdot$nm$^6$)
in order to account for the combined uncertainty in {\it all} the
$C_n$'s. In the absence of weak relativistic spin-interactions the
projection $m_f=m_{fa}+ m_{fb}$ and the rotational angular momentum of
the atoms about the center of mass $\ell$ are good quantum numbers for
the collision event. In this work we limit ourselves to $\ell=0$
collisions which, at least in the absence of $\ell > 0$ low-energy
resonances, give the dominant contribution to the interaction in the
ultracold regime.
\newline \indent
Two key measurements on $^{40}$K-$^{87}$Rb mixtures provide the best
range of input parameters to our model available to date. First, the
elastic cross-section for $(9/2,9/2)$+$(2,2)$ has been directly
measured by inter-species thermalization~\cite{roati02prl} and by
damping of coupled K-Rb dipole oscillations~\cite{modugno02science}.
Combining the systematic uncertainties in these two determinations we
estimate $a_t=a_{\frac{9}{2}\frac{9}{2},22}=-410_{-80}^{+80}a_0$
(1~$a_0$~=~0.0529177~nm).  Mass scaling of this value to
$^{41}$K-$^{87}$Rb allows us to strongly tighten the upper bound on the
triplet scattering length of~Ref.~\cite{ferrari}, leading to
$a_t(41-87)$~=~156$^{+10}_{-5}$ and to reduce the original uncertainty
in the number of triplet bound states to $N_b^t(41-87)=31\pm1$.
\newline \indent
Second, in order to determine $a_s$ we measure the $ ( 9/2, 9/2 ) $+
$(2, 1 )$ atom-loss rate in a nondegenerate sample. We prepare a doubly
spin-polarized sample in the magnetic trap containing about 10$^4$ K
atoms and 10$^5$ Rb atoms at temperatures in the range $(300-900)$~nK,
and then transfer about 30\% of the Rb sample in the $(2, 1)$ state
using a fast radio-frequency sweep. After a short rethermalization
period we measure the time-evolution of the number of K atoms, which
decays through inelastic collisions with Rb according to the rate
equation $\dot{n}^{K}(t)=-G^{inel} n^{K}_{9/2,9/2}(t)\,
n^{Rb}_{2,1}(t)$.
We find $G^{inel}$~=~5.0(2.5)$\times
10^{-12}$~cm$^3$~s$^{-1}$, where the uncertainty is dominated by that on
the atom numbers. In this work we make a conservative choice and adopt
only the order of magnitude $G = (10^{-12}-10^{-11})$~cm$^3$~s$^{-1}$.
\begin{figure}
\vspace*{10.5cm}
\includegraphics{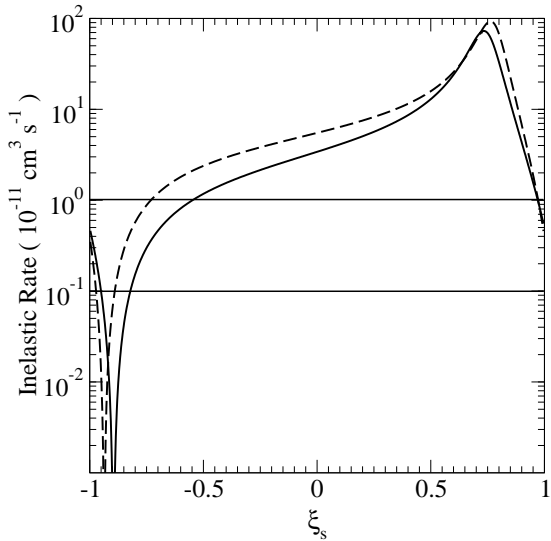} \caption{The numerically calculated
inelastic rate $G^{inel}$ for $^{40}$K$(9/2,9/2)$+$^{87}$Rb$(2, 1 )$
collisions at temperature $T=500$~nK as a function of the singlet
``scattering phase'' (see Text) for $a_t$~=~-330~$a_0$ (full line) and
$a_t$~=~-490~$a_0$ (dashed line). The horizontal lines correspond to the bounds
determined experimentally.  }
\label{inel_meas}
\end{figure}
\newline \indent
Figure~\ref{inel_meas} shows $G^{inel}$ numerically calculated as a
function of $a_s$ varying $a_t$ within the bounds determined above.
The horizontal axis is labeled by a scaled ``scattering phase''
$\xi_{s} =(2/{\pi}) \arctan ( a_s / a_{sc} )$, where
$a_{sc}=\frac{1}{2} (2 \mu C_6)^{1/4} \approx 72a_0$ is a typical
length for a purely van der Waals potential, with $\mu$ the K-Rb
reduced mass. To be noted the strong suppression of the inelastic rate
for $a_s\sim a_t$.  Comparison with the experimental value leads either
to a very large (positive or negative) or to a negative singlet
scattering length $ a_s(40-87)=-185_{-225}^{+83}a_0$. Mass-scaling
with $N_b^s(41-87)=98\pm2 $ gives a singlet scattering length for the
bosons pair $a_s(41-87)=5.0_{-34}^{+23}a_0$ and, by compatibility
with the $a_s(41-87)$ of Ref.~\cite{ferrari}, rules out the range of
very large $|a_s|(40-87)$.
With these $a_s,a_t$ we can predict the zero-field scattering length
and Feshbach resonances of different magnetic polarization states. 
Let us first consider the boson-fermion pair $^{40}$K-$^{87}$Rb.
\newline \indent
The inverted hyperfine structure of the fermionic potassium, with the
$f=9/2$ lying below the $f=7/2$ level, leads to a peculiar and very
favorable feature of this pair, namely the stability under
spin-exchange collisions of any combination where {\it either} atom is
in its lowest Zeeman sublevel. We therefore find it important to
summarize in Tab.~\ref{tableone} the scattering properites of the
$(9/2,-9/2)$+$(1,m_{fb})$ states and of at least the most easily
experimentally realizable  $(9/2,m_{fa})$+$(1,1)$  magnetic states.
Note all of these hyperfine states have a relatively large ( i.e. $|a|
>a_{sc}$ ) and negative zero-field scattering length. This ensures that
the two clouds will remain miscible even in the ultracold
regime, see Ref.~\cite{molmer98prl}.
\begin{table}[h]
\begin{center}
\caption{Compendium of $^{40}$K-$^{87}$Rb scattering properties. The
zero-field $s$-wave elastic scattering length $a$ is shown for
different hyperfine sublevels. The uncertainty in the resonance
positions $B_0$ results from the uncertainty in the model parameters
quoted in the text. The resonance widths $\Delta$ are shown for the
nominal values $a_s=-185a_0$, $a_t=-410a_0$ and $C_6=4274$~a.u. of
the scattering lengths and of the van der Waals coefficient.  }
\label{tableone}
\vskip 12pt
\begin{ruledtabular}
\begin{tabular}{ccccc} $(f_a,m_{fa})$+$(f_b,m_{fb})$ & $a$($a_0$) & $B_0$(G)&  $-\Delta$(G)  \\ \hline \\
(9/2,$\pm$9/2)+(2,$\pm$2)  &  $-410_{-80}^{+80}$  &    & \\
(9/2,$\pm$9/2)+(1,$\pm$1)  & $-235_{-121}^{+92}  $ &  &  & \\
(9/2,7/2)+(1,1)  & $ -245_{-120}^{+92}   $ & $319.6_{+13}^{-17}$    & 0.8  \\
                 &                       &   $920.3_{+93}^{-93}$    & 0.05  \\
(9/2,-7/2)+(1,1)  & $ -323_{-105}^{+90}    $ & $522.9^{+40}_{-55}$ &  0.3 \\
  &  & $564.8^{+20}_{-24}$  &  0.1  \\
  &  & $646.3^{+35}_{-52}$  &  4.6  \\
  &  & $658.0^{+50}_{-58}$  &  0.08  \\
  &  & $753.1^{+50}_{-59}$  &  0.3  \\
  &  & $787.4^{+60}_{-65}$  &  0.9  \\
(9/2,-9/2)+(1,1)  & $ -336_{-102}^{+89} $ & $505.3^{+31}_{-49}$  & 0.03  \\
  &  & $547.1^{+28}_{-28}$   & 0.2  \\
  &  & $593.9^{+34}_{-44}$   & 4.0  \\
  &  & $741.2^{+60}_{-65}$   & 0.9  \\
  &  & $921.0^{+92}_{-94}$   & $ 10^{-4}$  \\
(9/2,-9/2)+(1,0)  & $ -279_{-114}^{+92}  $  & $ 477.0^{+35}_{-47}  $   & $0.1 $ \\
                 &                        & $ 590.3_{+26}^{-31} $ &  $3.9$  \\
\end{tabular}
\end{ruledtabular}
\end{center}
\end{table}
\newline \indent
A magnetic Feshbach resonance arises when a molecular bound-state
crosses an atomic threshold~\cite{laue02pra}. Across a Feshbach resonance the
scattering length $a$ presents a typical dispersive dependence on the
strength of the applied field $B$, $a(B)= \tilde{a} (1-
\frac{\Delta}{B-B_0} )$, where $B_0$ is the field
strength on resonance, $\Delta$ is the resonance width and $\tilde a$ is
the off-resonance background scattering length.  
The resonance parameters $B_0$ and $\Delta$ have been extracted from an
accurate close-coupled calculation on a fine $B$-grid and are shown in
Tab.~\ref{tableone}.
\newline \indent
Homonuclear collisions between alkali atoms in a magnetic field have
been studied in great detail. Only a relatively small amount of
Feshbach spectroscopy is often necessary as input to theoretical models
in order to gain an impressive power in determining the interaction
potential and in predicting the position of additional resonances
\cite{Cs,vankempen02prl}. The parameters $a_s,a_t$ and $C_n$'s of our
collision model can in principle be determined following a similar
procedure. At least three resonances between the ones shown in
Tab.~\ref{tableone} would be needed as input to the model.
Inclusion of the high-field $(9/2,7/2)$+$(1,1)$ resonance would improve
the procedure since this latter resonance is nearly purely
triplet in character with average electron spin $\langle
 S^2 \rangle\approx$1.95 and very sensitive to $C_6$, $\delta B_0 /
B_0$~=~-1.2~$\delta C_6/C_6$.
\newline \indent
The relatively strong and tunable interaction between different
magnetic polarization states opens up the possibility of realizing
Bose-Fermi mixtures where the fermions undergo a Cooper-pairing phase
transition to a superfluid state via exchange of density fluctuations
of the background BEC~\cite{bijlsma00pra,viverit02pra}. Due to
identical particle symmetrization, direct as well as boson-mediated
$s$-wave collisions between a pair of fermionic K atoms are allowed
only in a Fermi gas with at least two internal components, which we
denote as ``1'' and ``2'' in the following. We also let $a_{Rb}$ be the
intra-species scattering length for collisions of Rb atoms, $a_K$ the
intra-species scattering length for collisions of K in state ``1'' and
K in state ``2'' and $a_{KRb,i}$, with $i=1,2$, the inter-species
scattering lengths for collisions of Rb and K in state ``$i$''.
\newline \indent
The critical temperature $T_c$ for boson induced $s$-wave Cooper
pairing can be given an analytic expression valid in the limit of low
boson-densities and of a BEC larger than the fermionic cloud, see
Eq.~(13) of Ref.~\cite{viverit02pra}. As long as $a_{KRb,1} a_{KRb,2} >>
a_{Rb} a_{K}$, the critical temperature is independent of $a_{K}$, and
increases exponentially with $a_{Rb} / (k_F a_{KRb,2} a_{KRb,1}) $,
where $k_F$ is the Fermi vector. It must be noted however that as soon
as $k_F a_{KRb,i}^{2} \sim a_{Rb} $, the $i$-th component of K and Rb
will become dynamically unstable and
collapse~\cite{modugno02science,roth02pra}. As a consequence the
optimal condition for Cooper pairing are $a_{KRb,1} \sim a_{KRb,2}
\equiv a_{KRb}$ and $a_{KRb}^2 k_F \sim a_{Rb}$.
\newline \indent
From Tab.~\ref{tableone} a mixture composed of K$(9/2 ,-9/2)$, K$(9/2,
-7/2)$ and Rb$(1,1)$ seems to be the most appropriate in order to
fulfill this condition.  However, for the nominal values of the
zero-field scattering lengths between these components, we estimate
that at least a factor of four increase in the average trapping
frequency would be necessary with respect to our current experimental
setup. If the actual $a_{KRb,i}$ turned out to be smaller than the
nominal value, this factor would increase even further.
Fortunately, at least for some potential parameters, a pair of
resonances exist that allow simultaneous tuning of $a_{KRb,1}$ and
$a_{KRb,2}$ to a large value {\it at the same time}, making such a large
trapping frequency unnecessary.
\begin{figure}
\vspace*{10.5cm}
\includegraphics{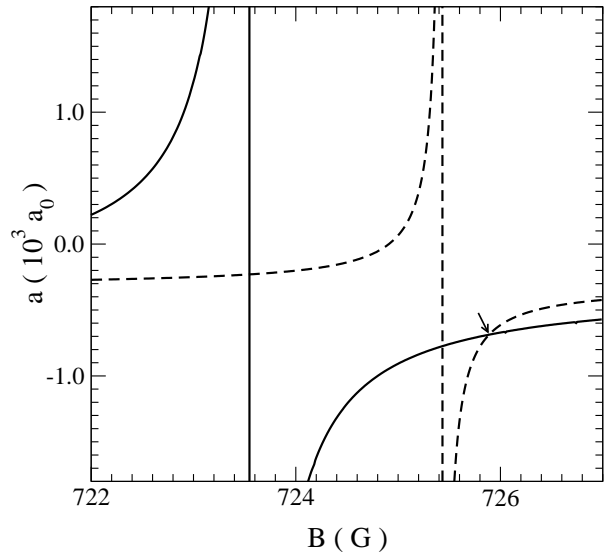} \caption{Field dependence of the
$^{40}$K$(9/2,-7/2)$+ $^{87}$Rb$(1,1)$ (full line) and
$^{40}$K$(9/2,-9/2)$+ $^{87}$Rb$(1,1)$ (dashed line) scattering lengths
for $a_s=-105 a_0$ and $a_t=-380 a_0$ and the nominal $C_6$. The
arrow shows the best working point for inducing the Cooper
pairing instability (see Text). }
\label{fig2}
\end{figure}
\newline \indent
A very favorable situation is shown in Fig.~\ref{fig2}, where for
$B\approx 725.89$~G we get $a_{KRb}\approx-687a_0$ . In order to have
this value stabilized to within 10~$a_0$ the magnetic field must be
stabilized to within $0.1$~G, which is a stringent yet possible to be met
requirement~\cite{vankempen02prl}.  The field at which $a_{KRb,1}=
a_{KRb,2}$ and the common value $a_{KRb}$ are extremely sensitive to
the details of the potential, and a conclusive word on what is the
largest achievable common $a_{KRb}$ is at this stage left to
forthcoming precision experiments on K-Rb Feshbach resonances.  As a
trend,  $|a_{s,t}|$ smaller than the nominal values tend to have a good
working field for inducing the Cooper pairing instability, for
$|a_{s,t}|$ larger than the nominal values (which however implies
larger zero-field $a$'s) such a field may not exist.
\begin{table}[h]
\begin{center}
\caption{Same than Tab.~\ref{tableone} but for $^{41}$K-$^{87}$Rb.
The
resonance widths $\Delta$ are shown for the nominal values
$a_s=5.0a_0$, $a_t=156a_0$ and $C_6=4274$~a.u. of
the scattering lengths and of the van der Waals coefficient.
In cases where the sign of $a$ is undetermined we have shown the nominal
value, the upper bound for $a<0$ and the lower bound for $a>0$.
}
\label{tabletwo}
\vskip 12pt
\begin{ruledtabular}
\begin{tabular}{cccc}   $(f_a,m_{fa})$+$(f_b,m_{fb})$  & $a$($a_0$) & $B_0$(G)& $-\Delta$(G)  \\ \hline \\
(2,2)+(2,2)  & $156_{-5}^{+10}$    &  &  \\
(2,2)+(2,1)  &    & $ 49.5_{-7}^{+5}$ & 6.5 \\
(1,1)+(1,1)  & 431,$ <-144$,$>249$     & $ 53.0^{+43}_{-53} $ & 30 \\
  &  & $ 87.5^{+37}_{-56} $    &  1.3 \\
  &  & $ 581.0_{+154}^{-166} $    & 77 \\
  &  & $ 740.5^{+136}_{-146} $   & 0.03 \\
(1,-1)+(1,-1)  & 431,$<-144  $,$>249  $ &     &   
\end{tabular}
\end{ruledtabular}
\end{center}
\end{table}
\newline \indent 
We now consider the
boson-boson pair $^{41}$K-$^{87}$Rb. 
We summarize in Tab.~\ref{tabletwo} the collision properties of the
states $(2,2)$+$(2,2)$ and $(1,1)$+$(1,1)$, which are collisionally
stable for any field, and of the state $(1,-1)$+$(1,-1)$ which is only
stable up to a threshold value $B^t \approx 143.45$ G, where the
$(2,-2)$+$(1,0)$ channel opens up and becomes the stable one. The
$a_{11,11}$ and $a_{1-1,1-1}$, which are equal at zero field, are
relatively large.
\newline \indent
The $(1,1)$+$(1,1)$ collision {\it always} presents at least two broad
and one narrow resonance below $10^3$~G, see Tab.~\ref{tabletwo} and 
Fig.~\ref{fig3}.
The analytical expression for the scattering length of a pair of
overlapping resonances with positions  $B_{0,\pm }$ and widths  $\Delta_{\pm }$
\bea
\frac{a(B)}{\tilde a}=1- \frac{(B-B_{0,+})\Delta_-+(B-B_{0,-})\Delta_+} 
{(B-B_{0,+})(B-B_{0,-})}, \nonumber 
\eea
has been used to fit both low-field resonances in the table.
The lowest-field resonance exists only if $a_{11,11}>0$, for
$a_{11,11}<0$ it turns into a real bound-state below the
$(1,1)$+$(1,1)$ atomic threshold and cannot be observed in a standard
scattering experiment.
\begin{figure}
\vspace*{10.5cm}
\includegraphics{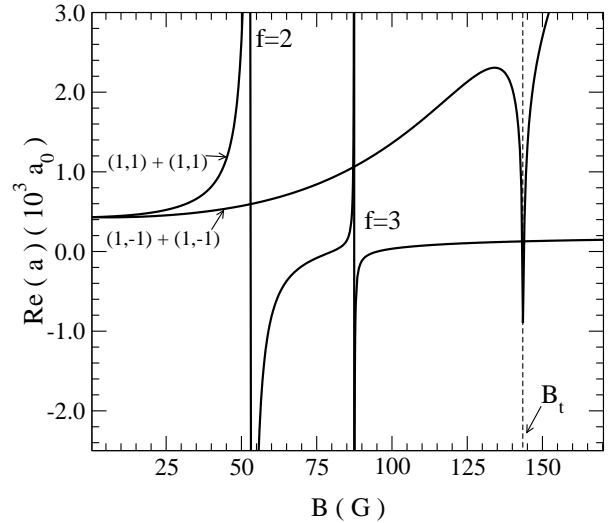} \caption{Field dependence of the purely real $a_{11,11}$
and of the real part ${\cal R}a_{1-1,1-1}$ for $^{41}$K-$^{87}$Rb
collisions and the nominal $a_s$, $a_t$, $C_6$. Approximate
zero-field quantum numbers $f$ , with ${\vec f}={\vec f}_a + {\vec
f}_b$, are assigned to the resonances. Near the $(2,-2)$+$(1,0)$
channel threshold $B_t$ (vertical line) ${\cal R}a_{1-1,1-1}$ changes
dramatically.
 }
\label{fig3}
\end{figure}
\newline \indent
The $(1,-1)$+$(1,-1)$ collision presents a feature near the threshold
value $B^t$. In this situation virtual excitations to the
$(2,-2)$+$(1,0)$ channel can strongly modify the real part of the
scattering length (see Fig.~\ref{fig3}), defined as ${\cal
R}a_{1-1,1-1} = -{\cal R} \lim_{k\to 0} S_{1-1,1-1}/(2ik)$, with $S$
the scattering matrix and $k$ the wave vector of the incoming atoms.
Such fluctuations-induced feature is always present in our range of
parameters if $a_{1-1,1-1}>0$ but turns out to be important for
$a_{1-1,1-1}<0$ only for very large $|a_{1-1,1-1}|$.  To be noted the
cusp at $B=B_t$ that follows from the analytical properties of the
scattering matrix near the opening of a new threshold~\cite{newton}.
\newline \indent
Finally, a field induced resonance in the $(2,2)$+$(2,1)$ inelastic
loss-rate has been included in Tab.~\ref{tabletwo} since it may be
important in order to constrain the collision model from
Feshbach-resonance spectroscopy carried out on the boson-boson
isotope. The width $\Delta$ shown in the table is in this case the
inelastic one, defined through the expression of the total scattering
eigenphase across resonance, $\delta(B) = \delta_0 -
\arctan{\frac{\Delta^{inel} }{ ( B - B_0 ) } }$, with $\delta_0$ the
background scattering phase. Moreover, near this resonance we find a
dramatic reduction of the $(2,2)$+$(2,1)$ inelastic decay rate down to
$G\approx 10^{-15}$~cm$^3$~s$^{-1}$, a value that ensures the
collisional stability of a two-species $(2,2)$+$(2,1)$ BEC.
\newline \indent
An interesting application of our results is to the production of polar
molecules via photoassociation, as proposed in Ref.~\cite{paul}.  First
step in this scheme is a Mott-insulator transition of a double K-Rb BEC
into a state where only pairs of molecules are loaded into the cells of
an optical potential. Whether K-K and Rb-Rb or K-Rb pairs are loaded
together depends upon the ratio of the interaction strength between all
pairs.  Using the low-field resonance shown in Fig.~\ref{fig3} one can
clearly vary the interaction from strongly attractive, thus favoring
formation of K-Rb pairs, to strongly repulsive, thus favoring K-K and
Rb-Rb pairs.  For instance, the condition $a_{KRb(11,11)}=0.5 \sqrt{
a_{K(11,11)} a_{Rb(11,11)}}$ required in Ref.~\cite{paul} in order to
load a single K and a single Rb per cell can easily be reached.  A
Raman photoassociation pulse or a direct microwave pulse can then be
used to associate atoms into molecules at each lattice cell.
In alternative, a time-dependent magnetic field has also been proposed
for near-unity conversion of atoms into molecules, see Ref.~\cite{mies00pra}.
This technique can of course be used for both the isotopic pairs
investigated in this work.
\newline \indent
We are grateful to P. S. Julienne, E. Tiesinga, and C. J. Williams for
stimulating discussions and for making available the ground-state
collisions code. We aknowledge useful discussions with F. Riboli and L.
Viverit. This work was supported by MIUR under a PRIN project, by ECC
under the Contract HPRICT1999-00111, and by INFM, PRA "Photonmatter".


\begin{thebibliography}{99}
\bibitem{pot} The more recent singlet $^1\Sigma$ and triplet $^3\Sigma$
potentials of Ref.~\cite{russeau} are adopted in this work.
\end{thebibliography}
\end{document}